# Guided Waves on Scalar and Tensorial Reactance Surfaces Modulated by Periodic Functions: a Circuital Approach.

Massimiliano Casaletti

***Abstract*—A technique to derive the propagation characteristics and field distributions of waves guided by scalar and tensorial reactance surfaces modulated by periodic or discrete Fourier spectrum functions in the propagation direction is presented. The method, based on an equivalent lumped circuits approach, can be seen as a generalization of the Oliner's method for the TM propagation on scalar sinusoidally modulated reactance surfaces. Numerical results are obtained for both surface wave and leaky wave solutions. The relevance of these studies to modulated metasurface antennas is discussed.**

***Index Terms*—**periodic surface, surface-waves, leaky-waves, metasurface antenna.

## I. Introduction

THE present study of guided waves on scalar and tensorial modulated reactive surfaces was motivated by recent developments in metasurface applications. Metasurfaces, described in terms of tensorial or scalar surface impedances, have been recently used in many applications like holographic and leaky-wave antennas, planar lenses, polarization convertors, orbital angular momentum communication or transformation optics [1]-[3]. By choosing appropriate modulated surface impedances, it is possible to control the propagation of Surface Waves (SW) along a surface or to obtain the transition from SW to leaky wave (LW) modes in order to realize antennas [5]-[12]

Most of these works are based on the propagation properties of waves over scalar sinusoidally modulated reactance surfaces [4]. Scalar metasurface antennas can produce general polarized beams [9]. However, they are limited by the fact that the direction of the radiating aperture field (or the equivalent surface current) is dictated by the source [9]. This latter aspect limits the number of possible aperture field distributions that can be implemented. Recently, tensorial modulated metasurfaces were successfully used in antenna design [5],[10]-[12]. On one hand, the additional degrees of freedom offered by tensorial metasurfaces could be used to overcome the limits of scalar solutions. On the other hand, these designs need modulated boundary conditions that cannot be analyzed using the formulation in [4].

The objective of this work is to extend the work done by Oliner and Hessel [4] to the analysis of propagation characteristics and field distributions of waves guided by scalar and tensorial reactance surfaces modulated by general periodic functions or by functions presenting a discrete Fourier spectrum.

This paper is structured as follows. Section II summarize the basic properties of propagation over periodic surfaces and introduces the formulation for the periodic scalar case. Approximated formulas are presented and applied to some examples of modulation. Section III extends the technique to the tensorial case. Periodic and non-periodic modulations having a discrete Fourier spectrum are considered. Finally, some considerations on transparent impedance modulations are presented in Section IV. Conclusions are drawn in Section V.

M. Casaletti is with the Sorbonne Universités, UPMC Univ. Paris 06, UR2, L2E, F-75005, Paris, France (e-mail: massimiliano.casaletti@upmc.fr).

## II. Scalar formulation

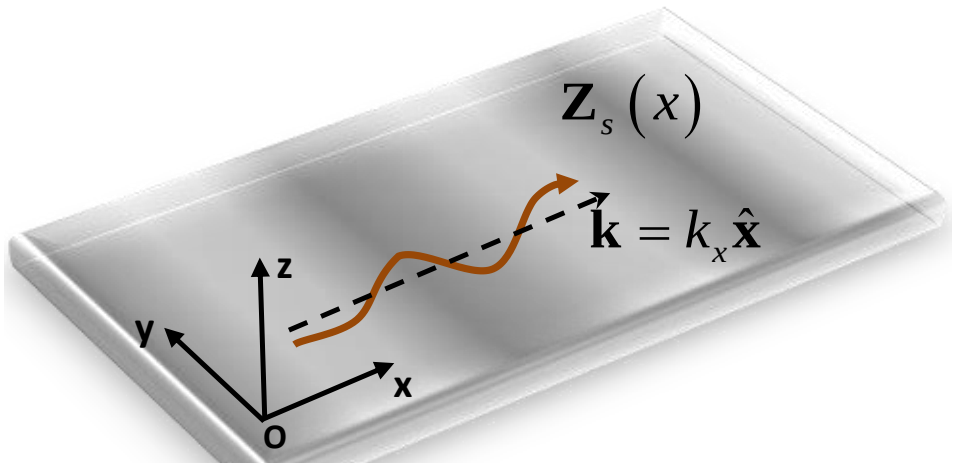

Figure 1. General geometry

This section presents a rigorous circuital model for the propagation characteristics and field distribution of TM and TE waves guided by planar surface reactance periodically modulated along the propagation direction.

The general geometry is shown in Fig. 1. The planar impedance surface is placed in the xy plane, while the waves guided by the surface are taken to propagate along the x direction.

A time dependence $e^{j\omega t}$, $\omega = 2\pi f$ being the angular frequency, is assumed and suppressed.

The surface impedance is supposed of the following general form:

$$Z_s(x) = j\bar{X}_s\left(1 + Mf(x)\right), \tag{1}$$

where $\bar{X}_s$ is the average value of surface reactance, $M < 1$ is the modulation index and $f$ is a periodic function of period $p$.

From Floquet's theorem the electromagnetic field can be decomposed in a discrete sum of modes (discrete spectrum of the wave operator) with $TE_z$ and $TM_z$ polarizations (Transverse Electric and Magnetic with respect to z-direction). The transverse fields can be expressed as

$$\begin{aligned}\mathbf{E}_t &= \sum_{n=-\infty}^{+\infty} V_n^{TM}(z)\mathbf{e}_n^{TM}(x,y) + \sum_{n=-\infty}^{+\infty} V_n^{TE}(z)\mathbf{e}_n^{TE}(x,y) \\ \mathbf{H}_t &= \sum_{n=-\infty}^{+\infty} I_n^{TM}(z)\mathbf{h}_n^{TM}(x,y) + \sum_{n=-\infty}^{+\infty} I_n^{TE}(z)\mathbf{h}_n^{TE}(x,y)\end{aligned}, \quad (2)$$

where $\mathbf{e}_n, \mathbf{h}_n$ are the electric and magnetic orthonormal modal vectors, $V_n$ and $I_n$ are the voltage and current amplitudes, respectively.

In rectangular coordinates the normalized tangent modal vectors are given by [13]

$$\begin{aligned}\mathbf{e}_n^{TM}(x,y) &= \bar{\mathbf{e}}^{TM}(x,y)e^{-jk_x^{(n)}x} = \hat{\mathbf{x}}\left(1/\sqrt{2\pi}\right)e^{-jk_x^{(n)}x} \\ \mathbf{h}_n^{TM}(x,y) &= \hat{\mathbf{y}}\left(1/\sqrt{2\pi}\right)e^{-jk_x^{(n)}x} \\ \mathbf{e}_n^{TE}(x,y) &= -\hat{\mathbf{y}}\left(1/\sqrt{2\pi}\right)e^{-jk_x^{(n)}x} \\ \mathbf{h}_n^{TE}(x,y) &= \hat{\mathbf{x}}\left(1/\sqrt{2\pi}\right)e^{-jk_x^{(n)}x}\end{aligned}, \quad (3)$$

where $k_x^{(n)} = k_x^{(0)} + n(2\pi/p)$ and they satisfy the following relations:

$$\begin{aligned}\left\langle \mathbf{e}_n^{\nu}, \mathbf{e}_m^{\kappa} \right\rangle &= \left\langle \mathbf{e}_n^{\nu}, -\hat{\mathbf{z}} \times \mathbf{h}_m^{\kappa} \right\rangle = \delta_{nm}\delta_{\nu\kappa} \\ \left\langle \mathbf{h}_n^{\nu}, \mathbf{h}_m^{\kappa} \right\rangle &= \left\langle \mathbf{h}_n^{\nu}, \hat{\mathbf{z}} \times \mathbf{e}_m^{\kappa} \right\rangle = \delta_{nm}\delta_{\nu\kappa}\end{aligned}. \quad (4)$$

where $\delta$ is the Kronecker delta, and $\nu, \kappa$ =TM/TE.

The voltage and current amplitudes have to satisfy the transmission line equation along the z-direction:

$$\begin{aligned}V_n^{\nu}(z) &= V_n^{\nu+}(z) + V_n^{\nu-}(z) \\ I_n^{\nu}(z) &= I_n^{\nu+}(z) + I_n^{\nu-}(z) = V_n^{\nu+}(z)/Z_n^{\nu} - V_n^{\nu-}(z)/Z_n^{\nu}\end{aligned}. \quad (5)$$

The original problem can be interpreted as shown in Fig. 2(a), where an infinite number of independent transmission line along the normal direction (one for each index $n$ and polarization state) are coupled together at the impedance surface.

The transmission line propagation constants and characteristic impedances are given by

$$\begin{aligned}k_z^{(n)} &= \sqrt{k^2 - k_x^{(n)2}} = \sqrt{k^2 - \left(k_x^{(0)} + n(2\pi/p)\right)^2} \\ Z_{(n)}^{TM} &= \zeta k_z^{(n)}/k \qquad Z_{(n)}^{TE} = \zeta k/k_z^{(n)}\end{aligned}, \quad (6)$$

where $k = \sqrt{\varepsilon\mu}$ is the wavenumber associated to the medium above the impedance and $\zeta$ the corresponding wave impedance.

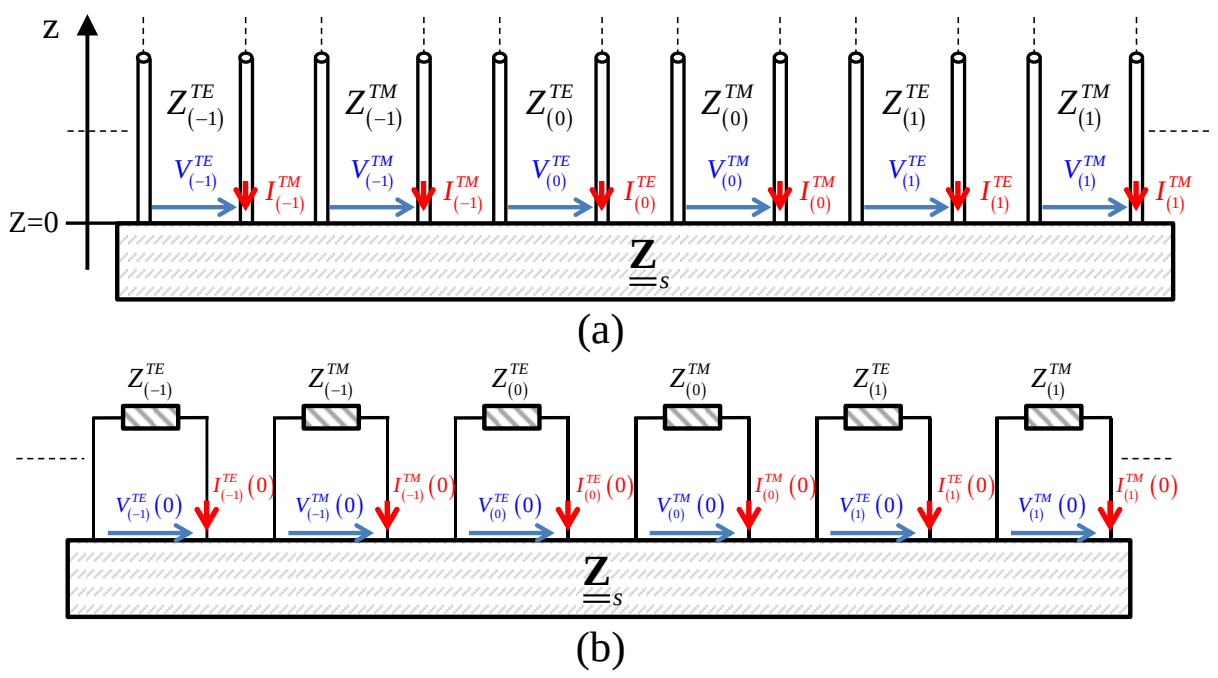

Figure 2: (a) Equivalent transmission line problem for $z$>0. (b) Equivalent circuit at $z$=0.

At the impedance level (z=0) the circuit can be simplified as in Fig.2(b), where the infinite transmission lines have been replaces by their corresponding line impedances.

Boundary condition (1) implies the following relation between electric and magnetic field:

$$\mathbf{E}_t(x) = Z_s(x)\hat{\mathbf{n}} \times \mathbf{H}_t \ . \quad (7)$$

Substituting the expansion (2) in (7) yields

$$\begin{aligned}&\sum_{n=-\infty}^{+\infty} V_n^{TM}(0)\mathbf{e}_n^{TM} + \sum_{n=-\infty}^{+\infty} V_n^{TE}(0)\mathbf{e}_n^{TE} \\ &= Z_s(x)\sum_{n=-\infty}^{+\infty} I_n^{TM}(0)\hat{\mathbf{n}} \times \mathbf{h}_n^{TM} + Z_s(x)\sum_{n=-\infty}^{+\infty} I_n^{TE}(0)\hat{\mathbf{n}} \times \mathbf{h}_n^{TE}\end{aligned}. \quad (8)$$

The periodic real function $f$ in (1) can be expressed as a Fourier series expansion

$$f(x) = \sum_{n=-\infty}^{\infty} c_n e^{-jn\frac{2\pi}{p}x} , \quad (9)$$

where $c_n^* = c_{-n}$. Thus, the modulated impedance can be rewritten as

$$Z_s(x) = Z_{self} + \sum_{m=-\infty}^{\infty} Z_{c,m} e^{-jm\frac{2\pi}{p}x} , \quad (10)$$

where the following quantities have been introduced:

$$\begin{aligned}Z_{self} &= j\bar{X}_s \\ Z_{c,n} &= j\bar{X}_s M c_n\end{aligned}. \quad (11)$$

Finally, using (10)-(11) in (8), testing with the TM and TE electric modal vector ($\mathbf{e}_n^{\nu}$), and using the normalization properties (4), leads to two independent families of scalar equations (one for each polarization):

$$\begin{aligned}V_n^{TM}(0) &= Z_{self} I_n^{TM}(0) + \sum_{\substack{m=-\infty \\ m\neq 0}}^{\infty} Z_{c,m} I_{n-m}^{TM}(0) \\ V_n^{TE}(0) &= Z_{self} I_n^{TE}(0) + \sum_{\substack{m=-\infty \\ m\neq 0}}^{\infty} Z_{c,m} I_{n-m}^{TE}(0)\end{aligned}. \quad (12)$$

As expected, from the scalar nature of (1) follows that TM and TE propagation problems can be treated separately.

Eq. (12) explains how a general modal circuit in the representation of Fig. 2(a) is coupled with all the other modes. Using circuit formalism, condition (12) states that in each modal circuit, the boundary condition is equivalent to the impedance $Z_{self}^{\nu}$ connected to a controlled voltage source $V_{g,n}^{\nu}$ defined as

$$V_{g,n}^{\nu} = \sum_{\substack{m=-\infty \\ m\neq 0}}^{\infty} Z_{c,m}^{\nu} I_{n-m}^{\nu}\left(0\right). \quad (13)$$

Fig. 3(a) summarizes the exposed equivalence; while Fig. 3(b) introduces an equivalent compact representation based on graph, where each node represents a mode and each arrow represents the interaction between different modes.

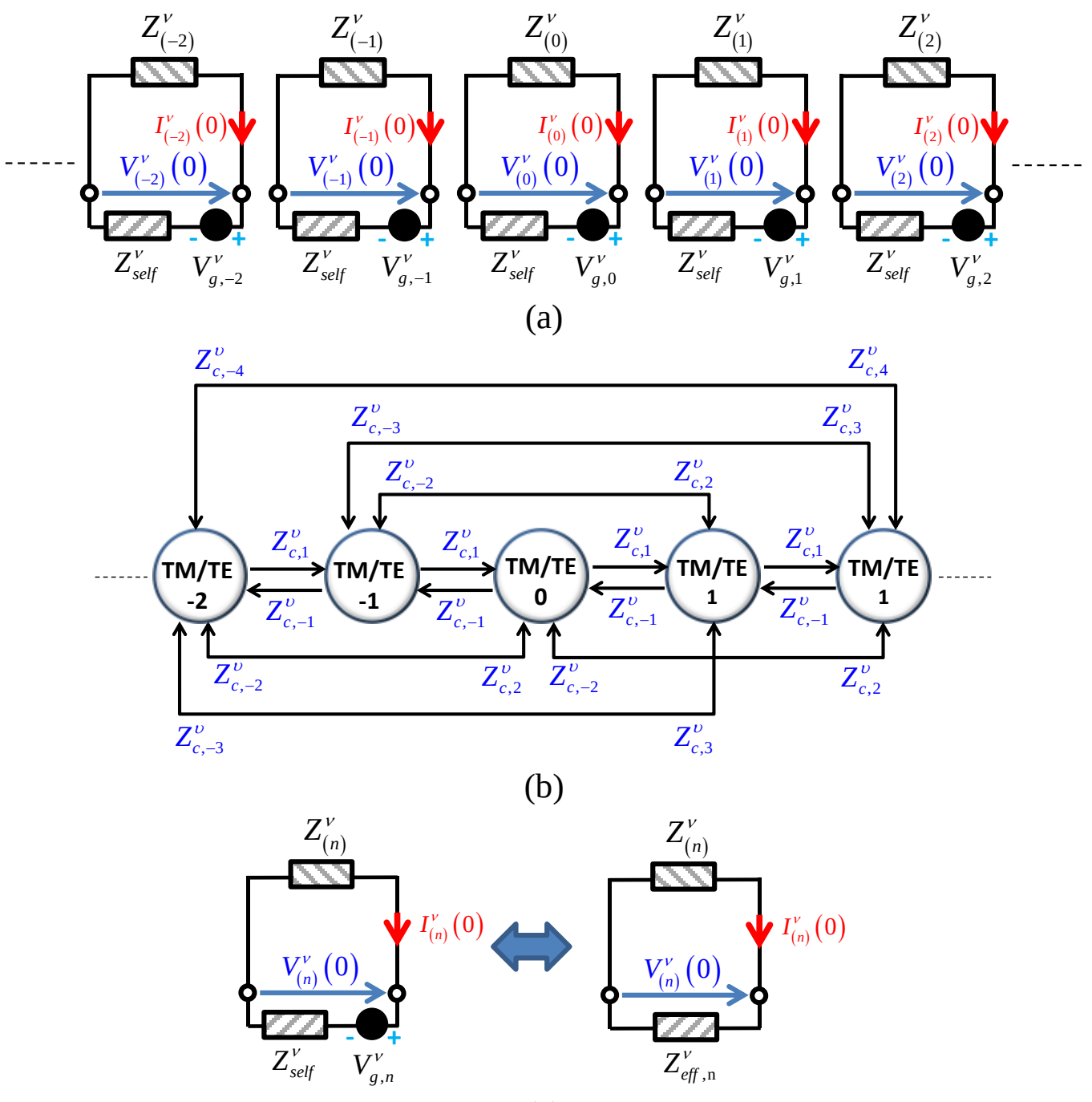

Figure 3. (a) Equivalent circuits at z=0. (b) Interactions graph for a general modulation. (c) Effective mode impedance.

The propagation constant can be found by imposing the resonance of any circuit *n* (reference mode), namely

$$Z_{eff,n}^{\nu}\left(k_x\right) + Z_{n}^{\nu}\left(k_x\right) = 0, \quad (14)$$

where the effective impedance $Z_{eff,n}^{\nu}$ (shown in Fig.3(c)) is defined as

$$Z_{eff,n}^{\nu} = Z_{self}^{\nu} + V_{g,n}^{\nu} / I_{n}^{\nu}. \quad (15)$$

The impedance (15) can be interpreted as the effective impedance of the homogenized problem for a particular mode. The current $I_{n-m}^{\nu}\left(0\right)$ in (13) depends on the infinite interactions between all the modes (jumps between a node to the others). For practical considerations the interactions has to be limited to a finite number of jumps. The order of accuracy of the truncated solution with respect the small parameter $M$ is given by $M^{k}$, where *k* is the number of interactions/jumps between the reference mode and the different nodes of the graph.

For most type of modulations it is difficult to found an analytical expression for $Z_{eff,n}^{\nu}$ valid for any order of accuracy. An alternative formulation can be obtained using (13) in (15), then imposing resonance condition (14) leading to

$$I_{n}^{\nu}\left(0\right)\left(Z_{n}^{\nu}\left(k_x\right) + Z_{self}^{\nu}\right) + \sum_{\substack{m=-\infty \\ m\neq 0}}^{\infty} Z_{c,m}^{\nu} I_{n-m}^{\nu}\left(0\right) = 0. \quad (16)$$

Equation (16) can be rewritten in matrix form as

$$\underline{\underline{\mathbf{M}}} \cdot \mathbf{I} = 0, \quad (17)$$

where $\mathbf{I} = \left[\ldots \;\; I_{n-2} \;\; I_{n-1} \;\; I_{n} \;\; I_{n+1} \;\; I_{n+2} \;\; \ldots\right]^{T}$,

$$\underline{\underline{\mathbf{M}}} = \begin{bmatrix} & & & \vdots & & & \\ & A_{n-2}^{\nu} & Z_{c,n-1} & Z_{c,n-2} & Z_{c,n-3} & Z_{c,n-4} & \\ & Z_{c,n+1} & A_{n-1}^{\nu} & Z_{c,n-1} & Z_{c,n-2} & Z_{c,n-3} & \\ \ldots & Z_{c,n+2} & Z_{c,n+1} & A_{n}^{\nu} & Z_{c,n-1} & Z_{c,n-2} & \ldots \\ & Z_{c,n+3} & Z_{c,n+2} & Z_{c,n+1} & A_{n+1}^{\nu} & Z_{c,n-1} & \\ & Z_{c,n+4} & Z_{c,n+3} & Z_{c,n+2} & Z_{c,n+1} & A_{n+2}^{\nu} & \\ & & & \vdots & & & \end{bmatrix}, \quad (18)$$

and $A_{n}^{\nu} = Z_{n}^{\nu} + Z_{self}^{\nu}$. The non-trivial solutions of eq.(17) can be found by imposing $\det \underline{\underline{\mathbf{M}}} = 0$.

1) *Sinusoidal modulation example*

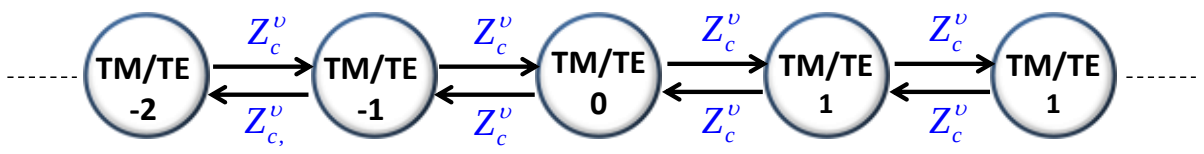

Figure 4. Interactions graph for a sinusoidal modulation.

Let us consider the simple case of a sinusoidal modulation:

$$f\left(x\right) = \cos\left(2\pi x / p\right). \quad (19)$$

Since $f$ is a pair function composed by two harmonics the cross-impedance defined in (11) reduces to

$$Z_{c,n}^{\nu} = \begin{cases} jM\bar{X}_s/2 = Z_c^{\nu} & n = 1,\text{-}1 \\ 0 & \text{otherwise} \end{cases}. \quad (20)$$

The nodes of the equivalent graph are connected only to the neighbors with the same interaction impedance as shown in Fig. 4. Using circuit theory, the effective impedance of a mode *n* can be written as an infinite continued fraction:

$$Z_{eff,n}^{\nu} = Z_{self}^{\nu} - \cfrac{Z_c^{\nu^2}}{Z_{self}^{\nu} + Z_{(n-1)}^{\nu} - \cfrac{Z_c^{\nu^2}}{Z_{self}^{\nu} + Z_{(n-2)}^{\nu} - \cfrac{Z_c^{\nu^2}}{Z_{self}^{\nu} + Z_{(n-3)}^{\nu} - \ldots}}} - \cfrac{Z_c^{\nu^2}}{Z_{self}^{\nu} + Z_{(n+1)}^{\nu} - \cfrac{Z_c^{\nu^2}}{Z_{self}^{\nu} + Z_{(n+2)}^{\nu} - \cfrac{Z_c^{\nu^2}}{Z_{self}^{\nu} + Z_{(n+3)}^{\nu} - \ldots}}} \quad (21)$$

However, as expected from the previous considerations, noticing that $Z_c^{\nu}$ is proportional to the small parameter M, the impedance value converges after some iterations. An example of TM and TE effective impedances for the mode *n*=0 is shown in Fig. 5. The modulation has been selected in order to impose $\sin(\theta_p)=0.3$, where $\theta_p$ is the pointing angle of the mode *n=-1*. Fig. 5(a) presents the TM and TE impedances as a function of the modulation indexes obtained using (21) with 3 and 51 modes, while Fig.5(b) shows the same quantities as a functions and of the average impedances. As can be seen the curves are superimposed, confirming that (21) can be safely truncated after few iterations. A practical rule is to include all the radiating modes (namely $\left|\mathrm{Re}\left\{k_x^{(n)}\right\}\right|<k_0$ ) in eq. (21). In the previous example this correspond to include the mode *n=-1*.

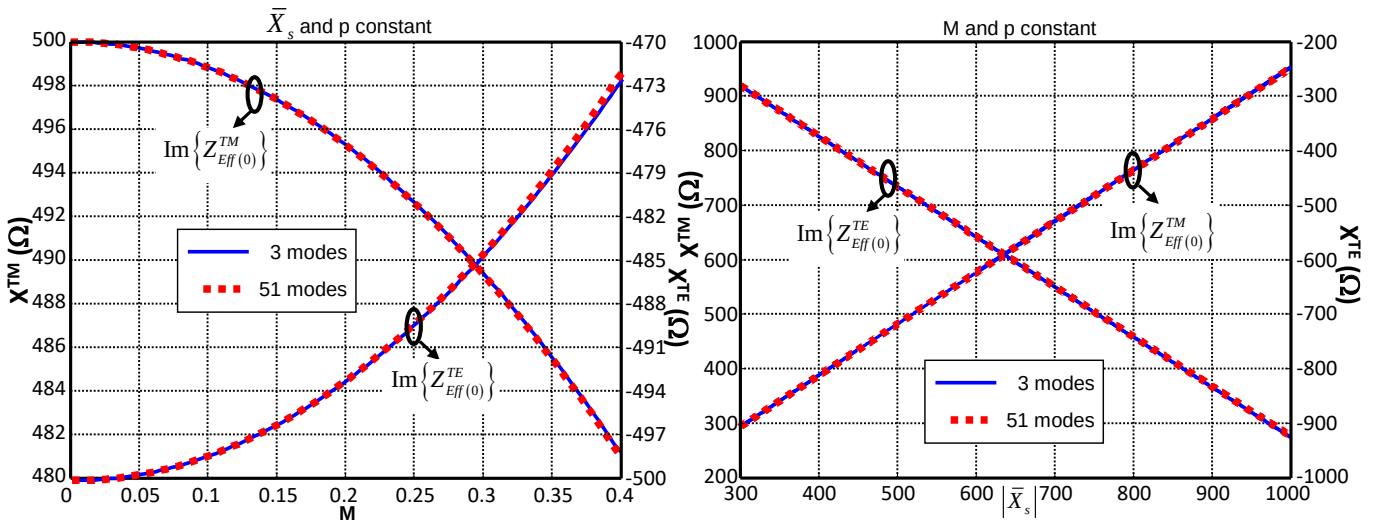

(a) (b)

Figure 5. Effective impedances for the mode *n*=0 at 30GHz, with a sinusoidal modulation defined by $\sin(\theta_p)=0.3$. (a)(c) TM/TE impedances as a function of the modulation index with $\bar{X}_s^{TM}=400$, $\bar{X}_s^{TE}=-400$; (b)(d) TM/TE impedances as a function of the average impedance with $M=0.2$.

## B. *Approximated solutions*

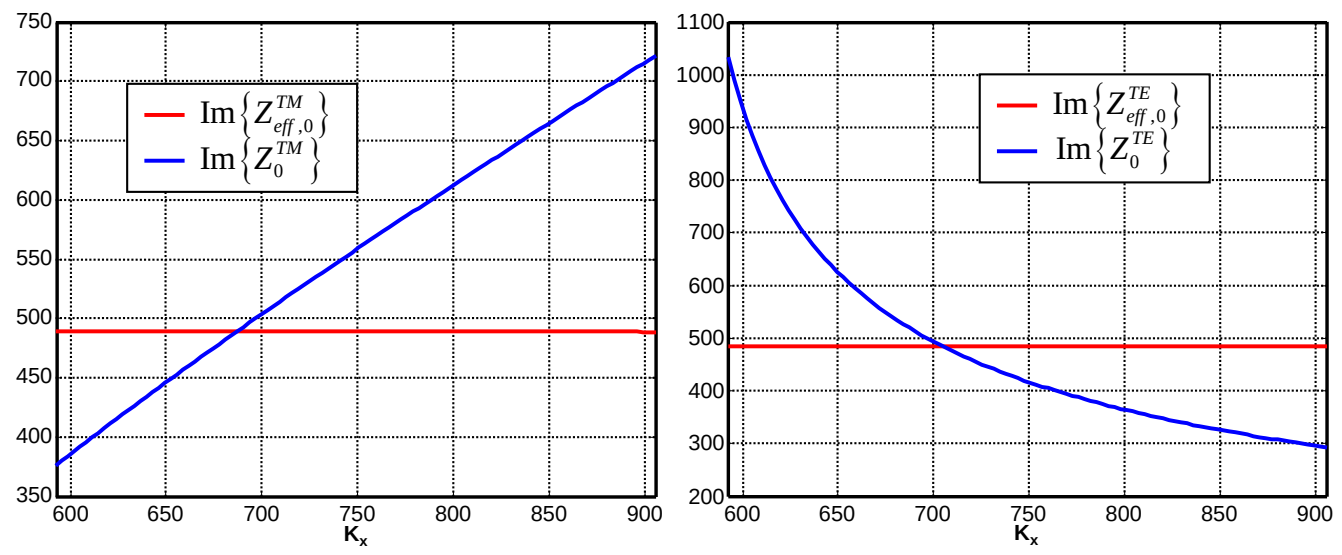

Figure 6: $Z_{eff,n}^{\nu}$ and $Z_n^{\nu}$ behaviors as a function of the wavenumber $k_x$ for a sinusoidal modulation. (a) TM polarization. (b) TE polarization.

The dispersion relation (14) can be approximated by noticing that the effective impedance $Z_{eff,n}^{\nu}$ is a slowing variating function of the spectral variable $k_x$ with respect to $Z_n^{\nu}$ (see Fig.6).

Using a zero order Taylor expansion for $Z_{eff,n}^{\nu}$ centered on the unmodulated impedance resonant wavenumber ( $\bar{k}_x=k_0\sqrt{1-\left(j\bar{X}_s/\zeta\right)^2}$ for TM polarization and $\bar{k}_x=k_0\sqrt{1-\left(\zeta/j\bar{X}_s\right)^2}$ for TE polarization) leads to

$$Z_{eff,n}^{\nu}\left(\bar{k}_x^{(n)}\right)+Z_n^{\nu}\left(k_x^{(n)}\right)=0\,, \tag{22}$$

where $\bar{k}_x^{(n)}=\bar{k}_x+n2\pi/p$.

Then, the wavenumber is easily obtained as

$$\begin{aligned} k_x^{TM(n)} &= k_0\sqrt{1-\left(Z_{eff,n}^{TM}\left(\bar{k}_x^{TM(n)}\right)/\zeta\right)^2} \\ k_x^{TE(n)} &= k_0\sqrt{1-\left(\zeta/Z_{eff,n}^{TE}\left(\bar{k}_x^{TE(n)}\right)\right)^2} \end{aligned}\,. \tag{23}$$

A further simplification can be obtained by decomposing the effective impedance in (22) as $Z_{eff,n}^{\nu}=Z_{self}^{\nu}+\Delta Z_{eff,n}^{\nu}$, and then expanding $Z_n^{\nu}$ at the first order

$$Z_n^{\nu}(k_x)=Z_n^{\nu}\left(\bar{k}_x\right)+\frac{dZ_n^{\nu}\left(\bar{k}_x\right)}{dk_x}\Delta k_x\,, \tag{24}$$

yielding

$$\Delta k_x=\Delta\beta_x-j\alpha_x=-\Delta Z_{eff,n}^{\nu}\left(\bar{k}_x\right)\Big/\frac{dZ_n^{\nu}\left(\bar{k}_x\right)}{dk_x}\,. \tag{25}$$

Equation (25) allows finding a direct relation between the modulating parameters and the variation of the wavenumber. Fig. 7 presents the real and imaginary part of the TM and TE propagation constant relative to the previously defined problem (Fig. 5) as a function of the modulation indexes and the average impedances. The blue dashed lines are the solutions of the exact dispersion relation (14) solved using a Padé approximant method. The red and green lines are obtained using (23) with 3 and 5 modes, respectively. Finally, the black curves are obtained with the approximation (25) using 3 modes.

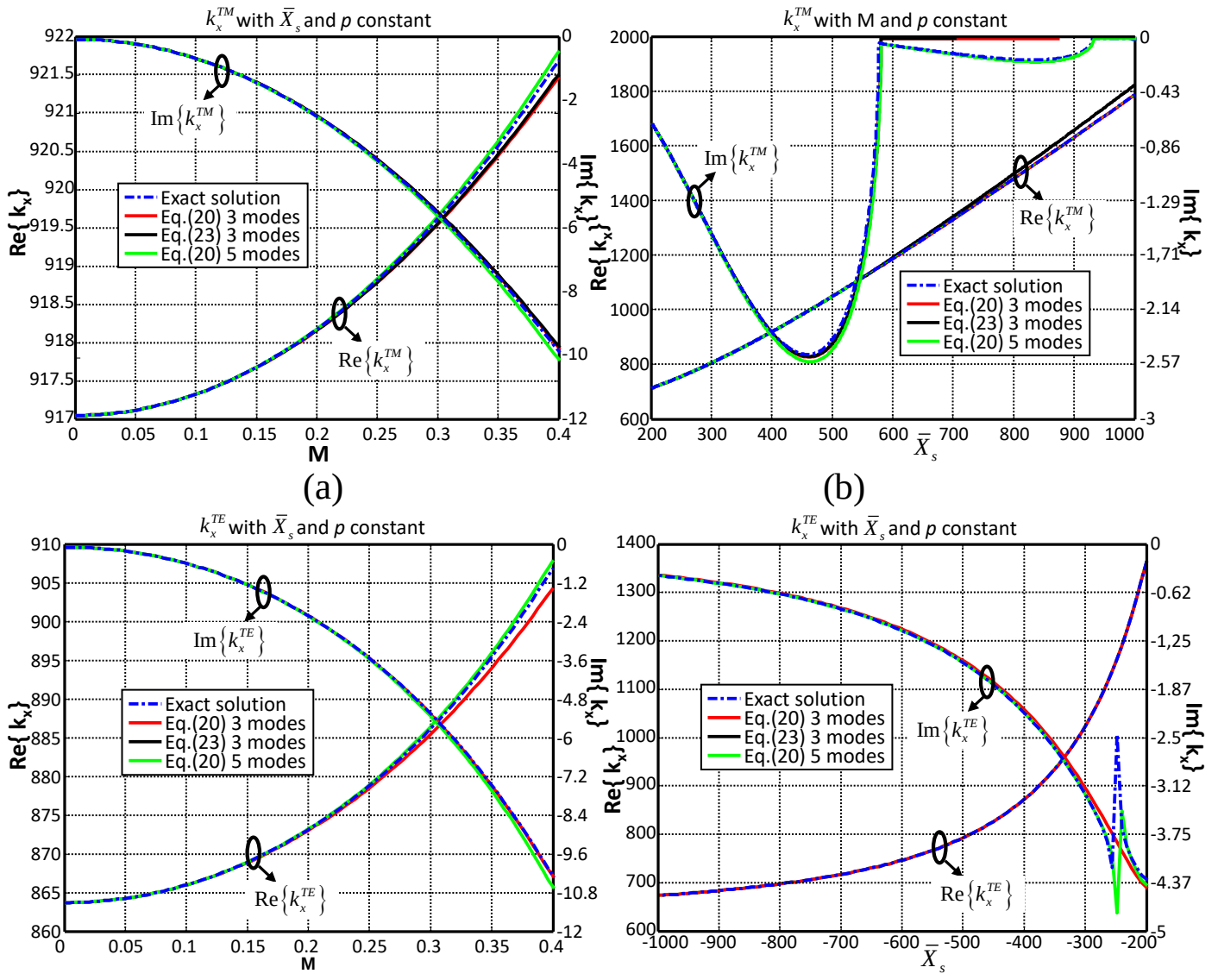

(a) (b)

(c) (d)

Figure 7: Real and imaginary part of the propagation constant as a function of the modulating index (a TM, b TE) and as a function of the average impedance ( b TM, c TE) obtained with three different methods.

## C. Application to scalar metasurface antennas

Scalar metasurface antennas [5]-[9] make uses of local scalar sinusoidal impedance modulations to convert a $TM_z$ cylindrical surface wave propagating from a feeder placed in the center to the periphery of the antenna into a leaky wave.

The local tangent problem is described by eqs. (1), (19). The feeder is designed in order to excite only the *n*=0 mode. This means that $I^{TM}_{(0)}$ corresponds to the magnetic field amplitude imposed by the source. The modulation period *p* is chosen in such a way that the *n*=-1 mode radiate to a desired direction $\theta_0$,

$$p = 2\pi / \left( k_x - k_0 \sin\theta_0 \right), \quad (26)$$

while all the other modes are evanescent along the z direction ($\left| k_x + n 2\pi/p \right| > k_0$ for $n \neq -1$).

The total tangent field on the metasurface antenna is given by the TM part of the expansion (2). Since there is just a propagating mode (n=-1), the far field radiated by the antenna can be expressed as the free space radiation of the equivalent electric or magnetic current relative to the *n*=-1 mode, namely:

$$\begin{aligned} \mathbf{J}^{(-1)}_{eq}(x,y) &= 2\hat{\mathbf{n}} \times \mathbf{h}^{TM}_{-1}(x,y) I^{TM}_{-1}(0) \\ \mathbf{M}^{(-1)}_{eq}(x,y) &= 2V^{TM}_{-1}(0) \mathbf{e}^{TM}_{-1}(x,y) \times \hat{\mathbf{n}} \end{aligned} \quad (27)$$

Using the proposed circuital approach, the current $I^{TM}_{(-1)}(0)$ can be written as a function of the incident field $I^{TM}_{(0)}$ as:

$$I^{TM}_{-1} = -\frac{\frac{M}{2}}{\gamma_{-1} - \left(\frac{M}{2}\right)^2 \dfrac{1}{\gamma_{-2} - \left(\frac{M}{2}\right)^2 \dfrac{1}{\gamma_{-3} - \left(\frac{M}{2}\right)^2 \dfrac{1}{\gamma_{-4} - \ldots}}}} I^{TM}_0, \quad (28)$$

where $\gamma_n = \left( 1 + Z^{TM}_n / jX_s \right)$.

Equivalently, the voltage is obtained as

$$V^{TM}_{-1} = Z^{TM}_{eff,-1} I^{TM}_{-1} = Z^{TM}_{-1}\left( k^{(-1)}_x \right) I^{TM}_{-1}. \quad (29)$$

Substituting (28)-(29) into (27) allows the calculation of the far-field radiated by the antenna. Moreover, equations (28)-(29) can be used to shape the equivalent current amplitude by changing the local modulation parameters.

## D. Non-sinusoidal modulation examples

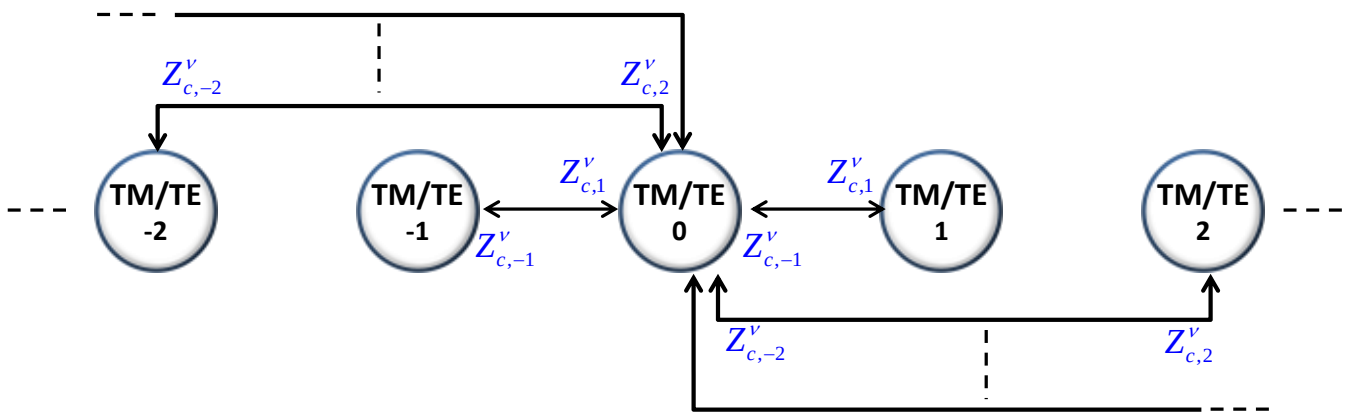

Figure 8. First order interactions graph for a general modulation.

In this section we apply the proposed formulation to the case of non-sinusoidal modulations. In particular we will presents numerical results for triangular and square modulations.

In section IIa-b it has been shown that the effective impedance can be approximated using a finite numbers of interactions. For simplicity, we limit our attention to the first order interaction (one jump between modes). The associated interaction graph is shown in Fig. 8.

Without loss of generality, let us consider the dominant mode (n=0). Using circuit theory its effective impedance can be written as

$$Z^{\nu}_{Eff(0)} = Z^{\nu}_{self} - \sum_{m=1}^{\infty} Z^{\nu}_{c,-m} Z^{\nu}_{c,m} \left( \frac{1}{Z^{\nu}_{self} + Z^{\nu}_{-m}} + \frac{1}{Z^{\nu}_{self} + Z^{\nu}_{m}} \right). \quad (30)$$

The associated dispersion relation is obtained using eq. (30) in (14). For practical reason, the series (30) has to be truncated. This can be done noticing that the convergence ratio of eq. (30) is directly related to the convergence of the Fourier expansion coefficients (9).

In the case that only one mode is radiating, using (30) in (25) and noticing that $Z^{\nu}_{c,-m} Z^{\nu}_{c,m} = -X_s^2 M^{\nu 2} \left| c_m \right|^2$ leads to the following expression of the attenuation constant:

$$\alpha^{\nu} = -\mathrm{Im}\left\{ k^{\nu}_x \right\} = \frac{2\pi M^2 \eta_\nu^{\,3} \left| c_{-1} \right|^2}{\lambda^2 \Omega} \frac{\sqrt{-\eta_\nu^{\,2}/\lambda^2 + 2\Omega/\lambda p - 1/p^2}}{2\Omega/\lambda p - 1/p^2} \quad (31)$$

where $\eta_{TM} = X_s/\zeta$, $\eta_{TE} = \zeta/\left| X_s \right|$, and $\Omega = \sqrt{1+\eta_\nu^{\,2}}$.

### 1) Triangular and square modulation

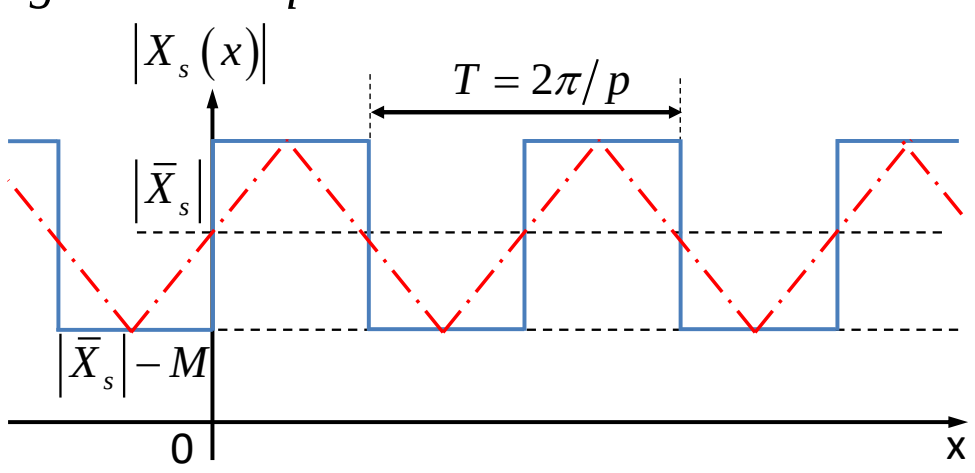

Figure 9 Triangular and square modulation profile.

Triangular and squared impedance modulations (see Figure 9) can be obtained using the following modulating functions:

$$f_{tri}(x) = \frac{8}{\pi^2} \sum_{k=0}^{\infty} \frac{(-1)^k}{(2k+1)^2} \sin\left( (2k+1)\frac{2\pi}{p} x \right), \quad (32)$$

$$f_{square}(x) = \frac{4}{\pi} \sum_{k=0}^{\infty} \frac{1}{(2k+1)} \sin\left( (2k+1)\frac{2\pi}{p} x \right) \quad (33)$$

From the fact that we deal with odd functions, it follows that

$$Z^{\nu}_{c,-m} = -Z^{\nu}_{c,m}. \quad (34)$$

Using (32) in (11) produces:

$$Z^{\upsilon}_{c,m} = \begin{cases} 8X_s^{TM} M^{TM} / (m\pi)^2 & \text{if } m \text{ odd} \\ 0 & \text{if } m \text{ even} \end{cases}, \quad (35)$$

while using (33) in (11) leads to

$$Z_{c,m}^{\upsilon} = \begin{cases} 4X_s^{TM} M^{TM} / m\pi \ \text{ if } m \text{ odd} \\ 0 \quad \text{if } m \text{ even} \end{cases}. \tag{36}$$

Fig. 10 presents a comparison between the propagation constant obtained with triangular, squared and sinusoidal modulations with $\sin\theta_0 = 0.3$ (only the mode $n$=-1 is radiating). The results have been obtained with the three different methods: numerically solving the dispersion relation, the approximation (23) and the expression (25). The curves obtained with the first two methods are almost superimposed, while the third one diverge from the exact solution for large values of M.

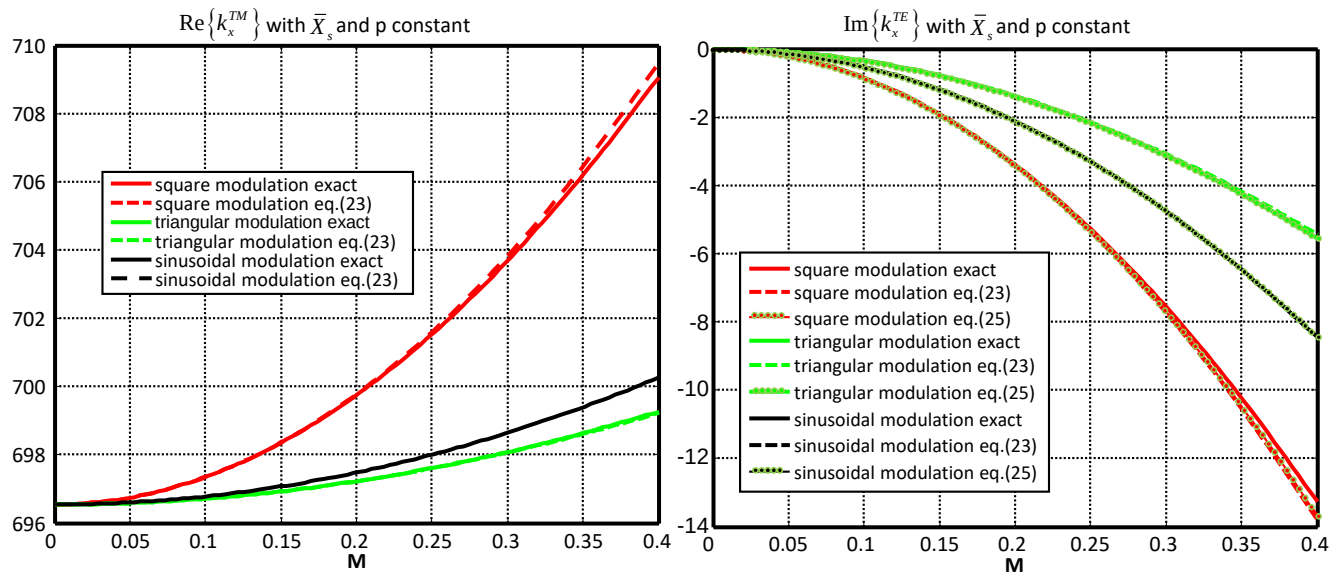

(a)

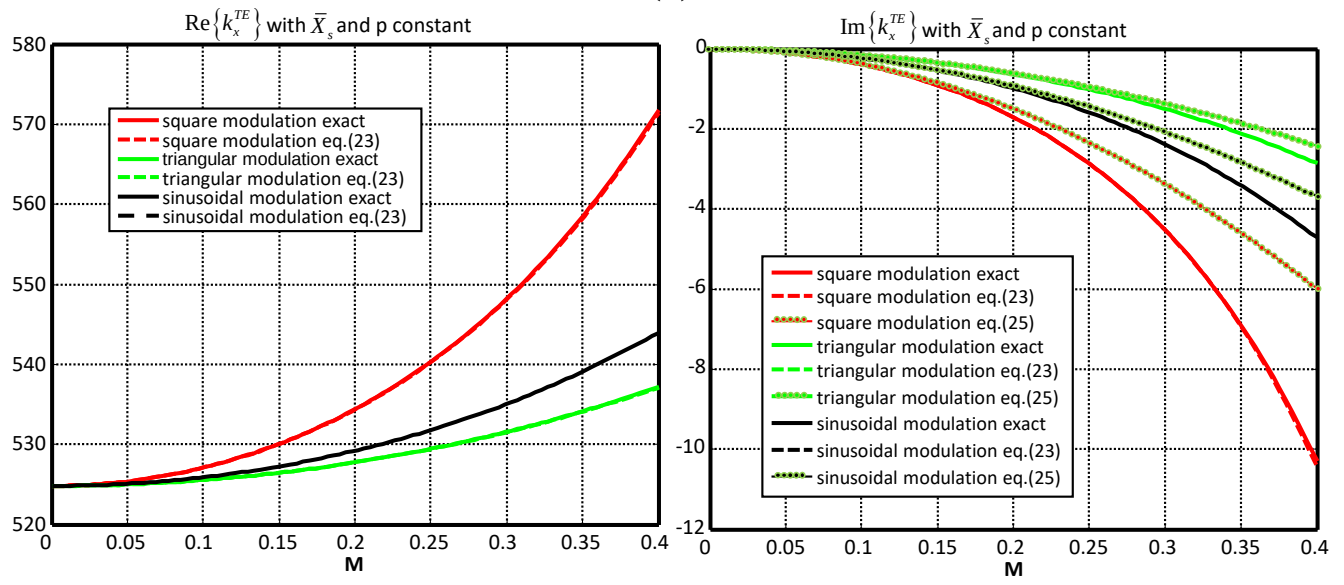

(b)

Figure 10 Real and imaginary part of the propagation constant as a function of the modulating index at 20GHz. (a) TM polarization with $\bar{X}_s = 500$; (b) TE polarization with $\bar{X}_s = -500$.

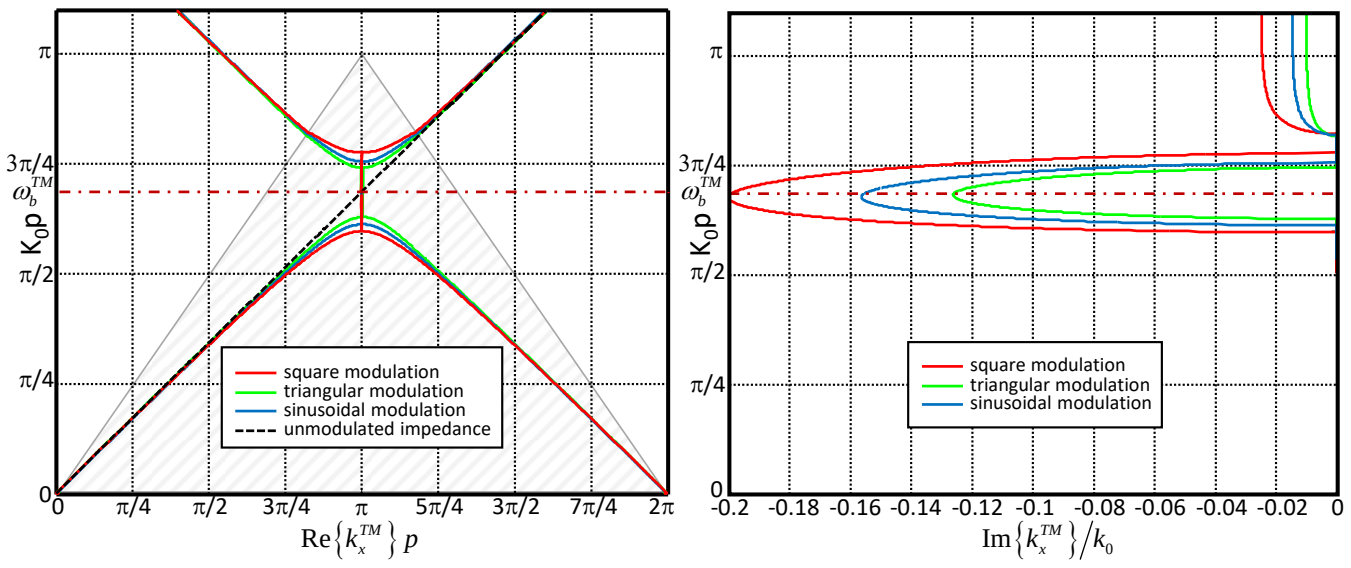

(a)

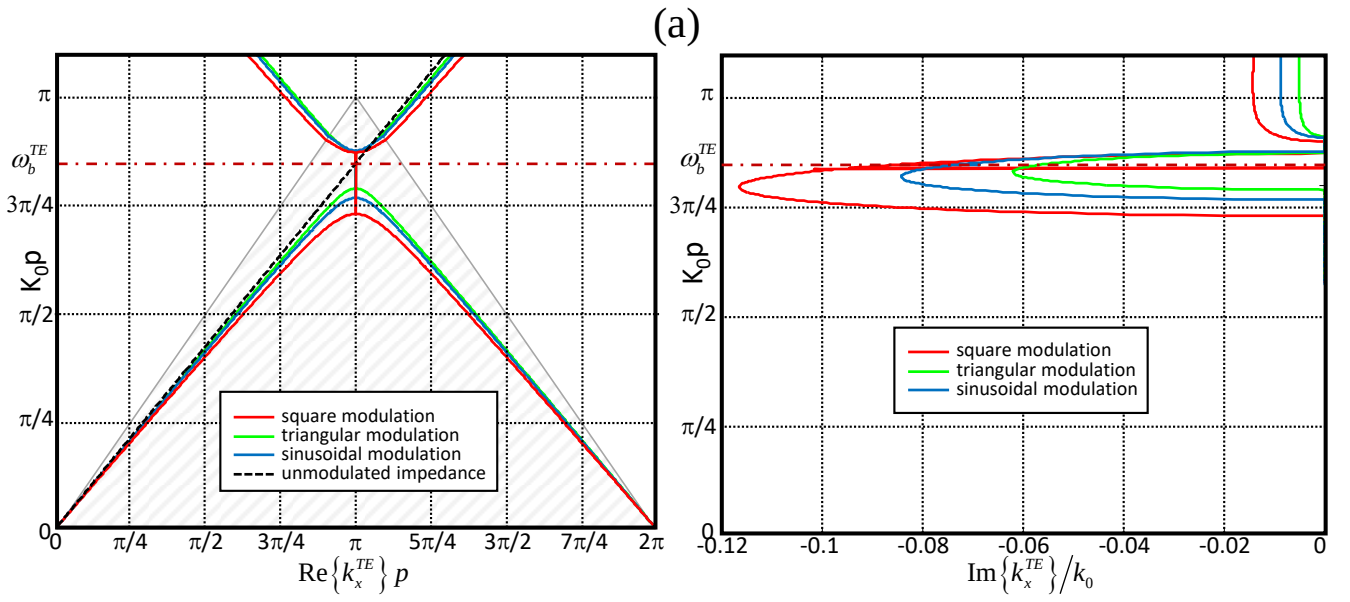

(b)

Figure 11 Dispersion diagrams (real part and imaginary part). (a) TM polarization; (b) TE polarization.

Since only the mode -1 is radiating, the attenuation constant is proportional to amplitude of the -1 harmonics ($|c_{-1}|$) of the different boundary conditions as predicted by (31). As can be seen, in this particular case the square modulation gives to the higher attenuation constant.

Fig. 11 shows the dispersion diagram (real part and imaginary part) for TM and TE polarizations. Surface wave solutions are defined in the dashed triangular area. The bandgap central position can be approximated as the intersection of the repeated unperturbed impedance solution, namely:

$$\begin{aligned} \omega_b^{TM} &= \pi \Big/ p_{TM}\sqrt{\varepsilon\mu\left(1+\left(\bar{X}_s^{TM}/\zeta\right)^2\right)} \\ \omega_b^{TE} &= \pi \Big/ p_{TE}\sqrt{\varepsilon\mu\left(1+\left(\zeta/\bar{X}_s^{TE}\right)^2\right)} \end{aligned}. \tag{37}$$

Thus, the bandgap position can be changed by acting on the periodicity ($p_\nu$) or the average impedance ($\bar{X}_s^\nu$). Fig.11(a) is relevant to the TM polarization with an average reactance $X_s = 400\Omega$, while Fig.11(b) reports the dispersion diagram for the TE polarization with average impedance $X_s = -600\Omega$ .As can be seen a larger bandgap is obtained around $k_x p = \pi$ with a square modulation.

## III. Tensorial Formulation

### A. *Canonical TM TE base periodic modulation*

Let us consider the tensorial formulation of the impedance boundary condition:

$$\mathbf{E}_t\big|_{z=0} = \underline{\underline{\mathbf{Z}}}_s \cdot \left(\hat{\mathbf{n}} \times \mathbf{H}_t\big|_{z=0}\right), \tag{38}$$

where the field and the impedance tensor are expressed in the TM/TE framework, namely

$$\begin{bmatrix} V_{TM} \\ V_{TE} \end{bmatrix} = \begin{bmatrix} Z_{TM,TM} & Z_{TM,TE} \\ Z_{TE,TM} & Z_{TE,TE} \end{bmatrix} \cdot \begin{bmatrix} I_{TM} \\ I_{TE} \end{bmatrix}. \tag{39}$$

Each component of the surface impedance tensor is supposed of the following general form

$$Z^{\nu,\kappa}(x) = j\bar{X}_s^{\nu,\kappa}\left(1 + M^{\nu,\kappa} f^{\nu,\kappa}(x)\right), \tag{40}$$

where $\nu,\kappa$ =TE/TM, $\bar{X}_s^{\nu,\kappa}$ is the average value of surface reactance, $M^{\nu,\kappa}$<1 is the modulation index and $f^{\nu,\kappa}$ is a periodic function with period $p_{\nu,\kappa}$.

If the periods $p_{\nu,\kappa}$ are multiple of a same period $p$, the problem is still periodic, thus it can be interpreted as composed by a discrete infinite number of independent transmission lines as for the scalar case (Fig.2). The only difference with respect to the scalar case is that lines with different polarizations (TE/TM) are coupled together.

In the following we will call this particular framework *the canonical basis* because eqs. (38)-(40) directly establish a relation between electrical quantities defined in different transmission lines (in general with different polarization state).

As for the scaler case, each periodic function $f^{\nu,\kappa}$ in (40) is expressed as Fourier series expansion

$$f^{\nu,w}\left(x\right)=\sum_{n=-\infty}^{\infty}c_{n}^{\nu,\kappa}e^{-jn\frac{2\pi}{p_{\nu,\kappa}}x}, \qquad (41)$$

then, using (41) in (40) leads to

$$Z^{\nu,\kappa}\left(x\right)=Z_{self}^{\nu,\kappa}+\sum_{m=-\infty}^{\infty}Z_{cross,\mathrm{m}}^{\nu,\kappa}e^{-jm\frac{2\pi}{p_{\nu,\kappa}}x}, \qquad (42)$$

where the following quantities have been introduced:

$$\begin{aligned}Z_{self}^{\nu,\kappa}&=j\bar{X}_{s}^{\nu,\kappa}\\ Z_{cross,\mathrm{m}}^{\nu,\kappa}&=j\bar{X}_{s}^{\nu,\kappa}M^{\nu,\kappa}c_{n}^{\nu,\kappa}\end{aligned} \qquad (43)$$

From eq.(40)-(43), it follows that the equivalent controlled voltage sources are defined as

$$\begin{aligned}V_{g(n)}^{TM}&=Z_{self}^{TM,TE}I_{n}^{TE}+\sum_{\substack{m=-\infty\\ m\neq n}}^{\infty}Z_{c,-m}^{TM,TE}I_{n+m}^{TE}+\sum_{\substack{m=-\infty\\ m\neq n}}^{\infty}Z_{c,-m}^{TM}I_{n+m}^{TM}\\ V_{g(n)}^{TE}&=Z_{self}^{TM,TE}I_{n}^{TM}+\sum_{\substack{m=-\infty\\ m\neq n}}^{\infty}Z_{c,-m}^{TM,TE}I_{n+m}^{TM}+\sum_{\substack{m=-\infty\\ m\neq n}}^{\infty}Z_{c,-m}^{TE}I_{n+m}^{TE}\end{aligned}, \qquad (44)$$

and the effective impedance can be defined as

$$\begin{aligned}Z_{eff,\mathrm{n}}^{TM}&=Z_{self}^{TM}+V_{g,\mathrm{n}}^{TM}/I_{n}^{TM}\\ Z_{eff,\mathrm{n}}^{TE}&=Z_{self}^{TE}+V_{g,\mathrm{n}}^{TE}/I_{n}^{TE}\end{aligned}. \qquad (45)$$

The dispersion can be calculated using (43)-(45) in (14). As for the scalar case, an alternative formulation can be obtained replacing (44) in (45) and imposing the resonance condition (14), leading to:

$$\begin{cases}I_{n}^{TM}\left(Z_{\mathrm{n}}^{TM}+Z_{self}^{TM}\right)+Z_{self}^{TM,TE}I_{n}^{TE}+\sum\limits_{\substack{m=-\infty\\ m\neq n}}^{\infty}Z_{c,-m}^{TM,TE}I_{n+m}^{TE}+\sum\limits_{\substack{m=-\infty\\ m\neq n}}^{\infty}Z_{c,-m}^{TM}I_{n+m}^{TM}=0\\ I_{n}^{TE}\left(Z_{\mathrm{n}}^{TE}+Z_{self}^{TE}\right)+Z_{self}^{TM,TE}I_{n}^{TM}+\sum\limits_{\substack{m=-\infty\\ m\neq n}}^{\infty}Z_{c,-m}^{TM,TE}I_{n+m}^{TM}+\sum\limits_{\substack{m=-\infty\\ m\neq n}}^{\infty}Z_{c,-m}^{TE}I_{n+m}^{TE}=0\end{cases}. \qquad (46)$$

Eq. (46) can be rewritten in matrix form as $\underline{\underline{\mathbf{M}}}\cdot\mathbf{I}=0$ where $\mathbf{I}=\left[\ldots I_{n-1}^{TM}\quad I_{n-1}^{TE}\quad I_{n}^{TM}\quad I_{n}^{TE}\quad I_{n+1}^{TM}\quad I_{n+1}^{TE}\ldots\right]^{T}$,

$$\underline{\underline{\mathbf{M}}}=\begin{bmatrix} & & & \vdots & & & & \\ \ldots & A_{n-1}^{TM} & Z_{self}^{TM,TE} & Z_{c,-1}^{TM,TE} & Z_{c,-1}^{TM} & Z_{c,-2}^{TM,TE} & Z_{c,-2}^{TM} & \\ & Z_{self}^{TM,TE} & A_{n-1}^{TE} & Z_{c,-1}^{TM,TE} & Z_{c,-1}^{TE} & Z_{c,-2}^{TM,TE} & Z_{c,-2}^{TE} & \\ & Z_{c,1}^{TM} & Z_{c,1}^{TM,TE} & A_{n}^{TM} & Z_{self}^{TM,TE} & Z_{c,-1}^{TM,TE} & Z_{c,-1}^{TM} & \\ \ldots & Z_{c,1}^{TM,TE} & Z_{c,1}^{TE} & Z_{self}^{TM,TE} & A_{n}^{TE} & Z_{c,-1}^{TM,TE} & Z_{c,-1}^{TE} & \ldots\\ & Z_{c,2}^{TM} & Z_{c,2}^{TM,TE} & Z_{c,1}^{TM} & Z_{c,1}^{TM,TE} & A_{n+1}^{TM} & Z_{self}^{TM,TE} & \\ & Z_{c,2}^{TM,TE} & Z_{c,2}^{TE} & Z_{c,1}^{TM,TE} & Z_{c,1}^{TE} & Z_{self}^{TM,TE} & A_{n+1}^{TE} & \\ & & & \vdots & & & & \end{bmatrix} \qquad (47)$$

and $A_{n}^{\nu}=Z_{\mathrm{n}}^{\nu}+Z_{self}^{\nu}$.

As done previously, the non-trivial solutions of eq.(47) are obtained imposing $\det\underline{\underline{\mathbf{M}}}=0$.

*1) Unmodulated impedance*

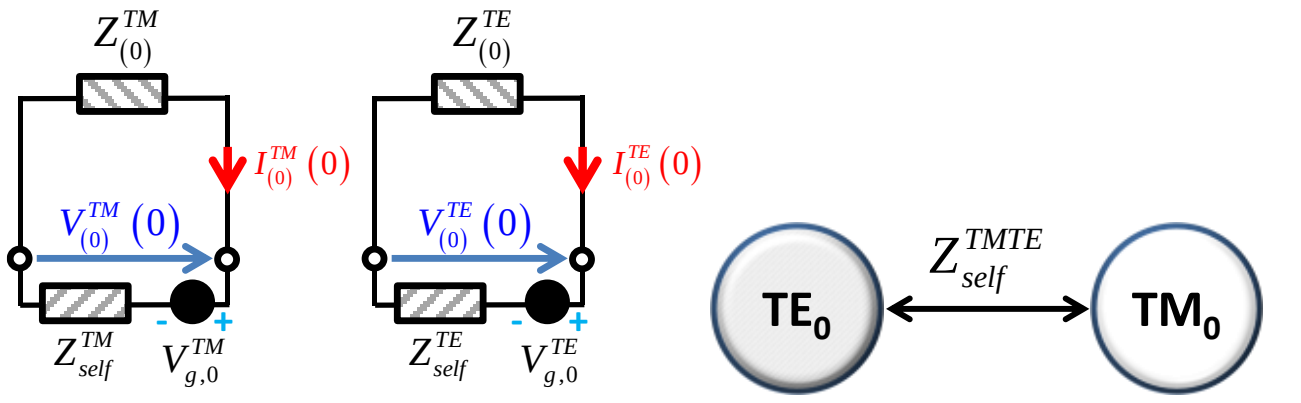

(a) (b)

Figure 12 (a) Equivalent circuit for an unmodulated tensor impedance condition. (b) Corresponding interaction graph.

The proposed circuital approach can be used to calculate the dispersion on the average impedance tensor in a natural way by selecting $M^{\nu,\kappa}=0$. The corresponding interaction graph and the equivalent circuit are shown in Fig. 12.

The effective impedance for the TM mode is given by

$$Z_{Eff}^{TM}=Z_{self}^{TM}-\frac{{Z_{self}^{TE,TM}}^{2}}{Z_{(0)}^{TE}+Z_{self}^{TE}} \qquad (48)$$

Then, the transmission line propagation constant is obtained as

$$\bar{k}_{z}=-j\bar{\alpha}_{z}=\frac{\left(\Lambda\pm\sqrt{\Lambda^{2}+4\bar{X}_{s}^{TE}\bar{X}_{s}^{TM}\zeta^{2}}\right)k_{0}}{2j\bar{X}_{s}^{TE}\zeta} \qquad (49)$$

where $\Lambda=\bar{X}_{s}^{TM}\bar{X}_{s}^{TE}-\left(\bar{X}_{s}^{TM,TE}\right)^{2}-\zeta^{2}$.

In agreement with [14], depending on the values of $\bar{X}_{s}^{TM}$, $\bar{X}_{s}^{TMTE}$ and $\bar{X}_{s}^{TE}$, four different cases are possible: 1) $\bar{X}_{s}^{TM}\bar{X}_{s}^{TE}>0$: $\bar{\alpha}_{z}$ only has a positive value, that means a SW type can propagate on the metasurface; 2) $\Lambda^{2}+4\bar{X}_{s}^{TE}\bar{X}_{s}^{TM}\zeta^{2}>0$: two values are possible for $\bar{\alpha}_{z}$ with the same sign as $\bar{X}_{s}^{TE}$ giving rise to either no SW or two SW with different propagation constants; 3) $\Lambda^{2}+4\bar{X}_{s}^{TE}\bar{X}_{s}^{TM}\zeta^{2}<0$: $\bar{\alpha}_{z}$ is complex, that does not correspond to any physical SW. 4) $\Lambda^{2}=4\bar{X}_{s}^{TE}\bar{X}_{s}^{TM}\zeta^{2}$ leading to $\bar{X}_{s}^{TMTE}=0$ and $\bar{X}_{s}^{TM}\bar{X}_{s}^{TE}=-\zeta^{2}$: $\bar{\alpha}_{z}$ only has a positive value corresponding to a TE and a TM waves propagating independently with the same propagation constant.

The ratio between TE and TM currents (and magnetic fields) amplitudes can be easily obtained as:

$$\left|\frac{I_{0}^{TE}}{I_{0}^{TM}}\right|=\left|\frac{Z_{self}^{TETM}}{Z_{0}^{TE}+Z_{self}^{TE}}\right|, \qquad (50)$$

this quantity will be useful later in order to understand the effect of each modulation index on the propagation constant.

*2) Sinusoidal, triangular and square modulations*

This section presents results for the propagation constant over a tensorial metasurface having the same kind of modulation for each component. Sinusoidal periodic modulation has been used in the literature to design tensorial metasurface antenna as in [10] and others. Three different modulating functions are investigated (sinusoidal, triangular and square modulation).

An average impedance with $\bar{X}_s^{TM} = 400$, $\bar{X}_s^{TE} = 200$, $\bar{X}_s^{TMTE} = 100$ at 20GHz is considered. Fig.13(a) shows the real and imaginary parts of $k_x$ for the three kinds of modulations (see legend) as a function of the modulating indexes. The continuous lines are relevant to variation of $M^{TM}$, dashed lines to variation of $M^{TE}$, and dotted lines to variation of $M^{TMTE}$. As for the scalar case, square modulation gives rise to the higher attenuation factor. Moreover, $M^{TM}$ produces higher variations of the propagation/attenuation constant. This latter aspect can be explained by noticing that from (49),(50) it follows that the current of the guided mode supported by the unmodulated impedance is dominated by the TM part. Thus, even in the perturbed boundary condition (modulation) the controlled voltage generator expression (44) is dominated by the TM terms. A dual behavior can be obtained changing the signs of $\bar{X}_s^{TM}$ and $\bar{X}_s^{TE}$ (dominant TE component) as shown in Fig. 13(b).

Fig.14 presents the dispersion diagrams for the two above mentioned metasurfaces. A hybrid modulation has also been considered (sinusoidal, triangular and square modulations for the TM, TE and TMTE components, respectively).

A simple expression for the position of the bandgap can be obtained approximating the unmodulated effective impedance (48) at $k_x = \pi/p$ as

$$Z_{Eff}^{TM}(\pi/p) \simeq \frac{j\bar{X}_s^{TM}\zeta - \bar{X}_s^{TM}\bar{X}_s^{TE} - j\bar{X}_s^{TMTE^2}}{\zeta + j\bar{X}_s^{TE}}, \tag{51}$$

then, the position of the bandgap is obtained as for the scalar case:

$$\omega_b = \pi \Big/ p_{TM}\sqrt{\varepsilon\mu\left(1-\left(Z_{Eff}^{TM}(\pi/p)/\zeta\right)^2\right)}. \tag{52}$$

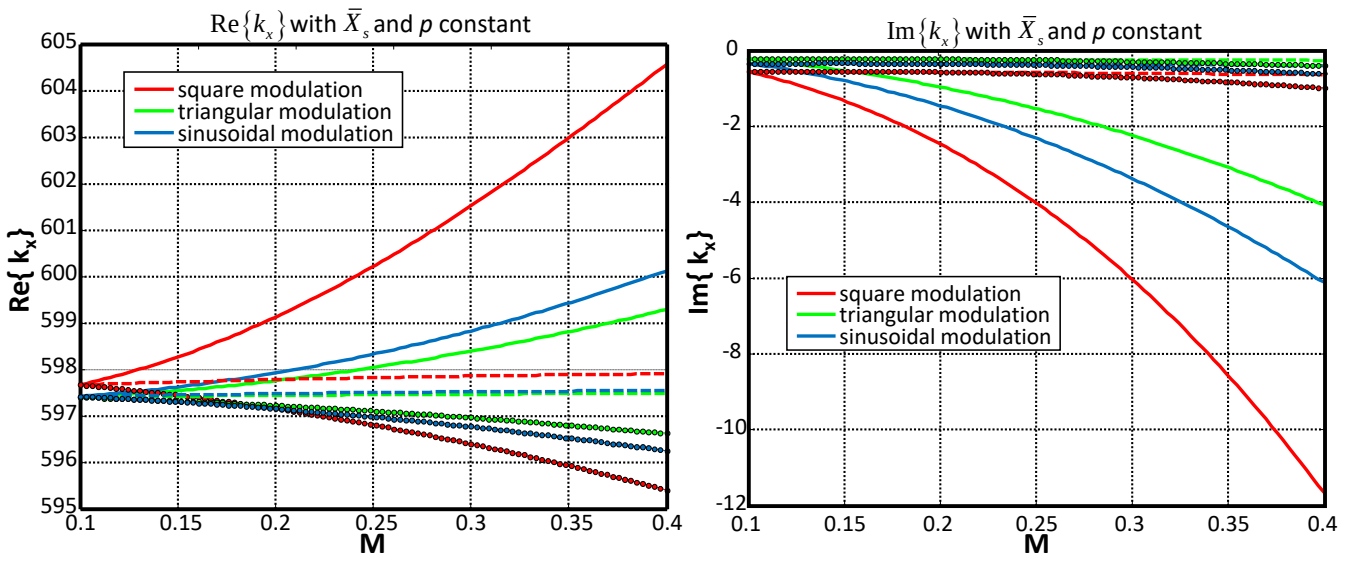

(a)

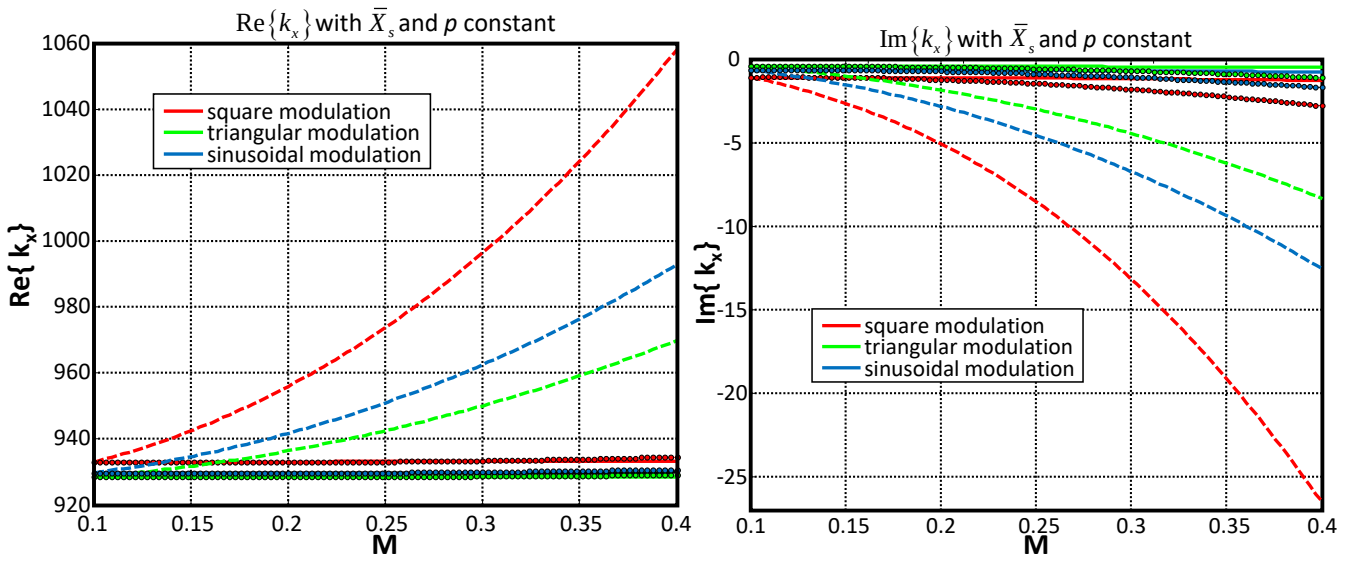

(b)

Figure 13. Real and imaginary part of $k_x$ as a function of the modulating indexes at 20Ghz for three kind of modulations. (a) $\bar{X}_s^{TM} = 400$, $\bar{X}_s^{TE} = 200$, $\bar{X}_s^{TMTE} = 100$. (b) $\bar{X}_s^{TM} = -400$, $\bar{X}_s^{TE} = -200$, $\bar{X}_s^{TMTE} = 100$.

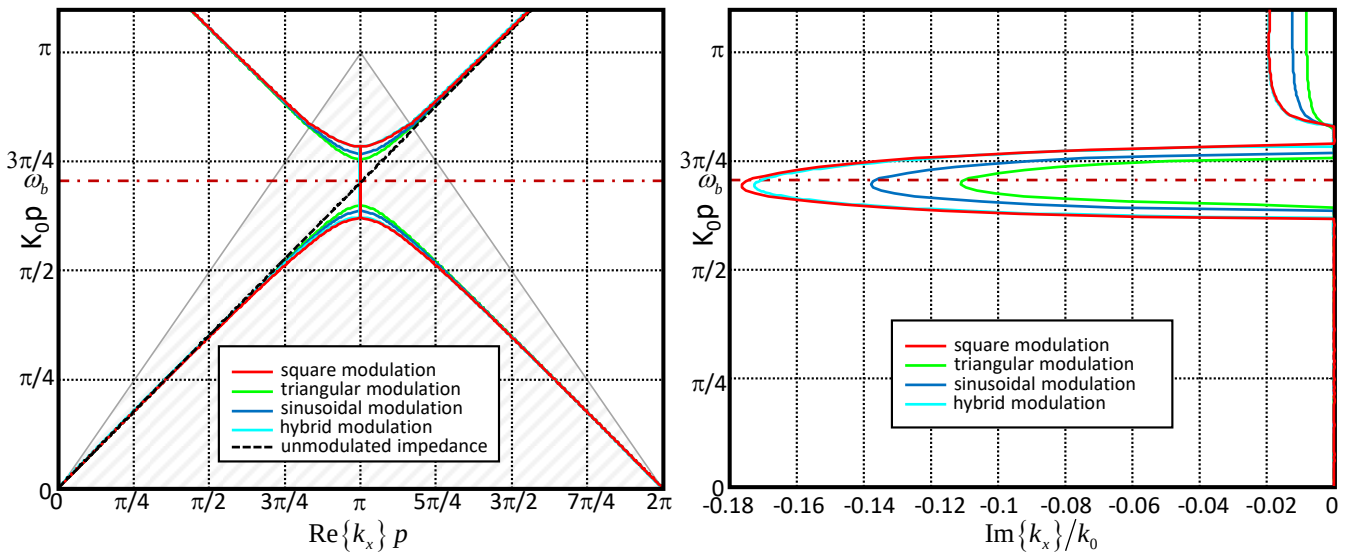

Figure 14; Dispersion diagram (real part and imaginary parts).

## B. *Canonical TM TE base modulation*

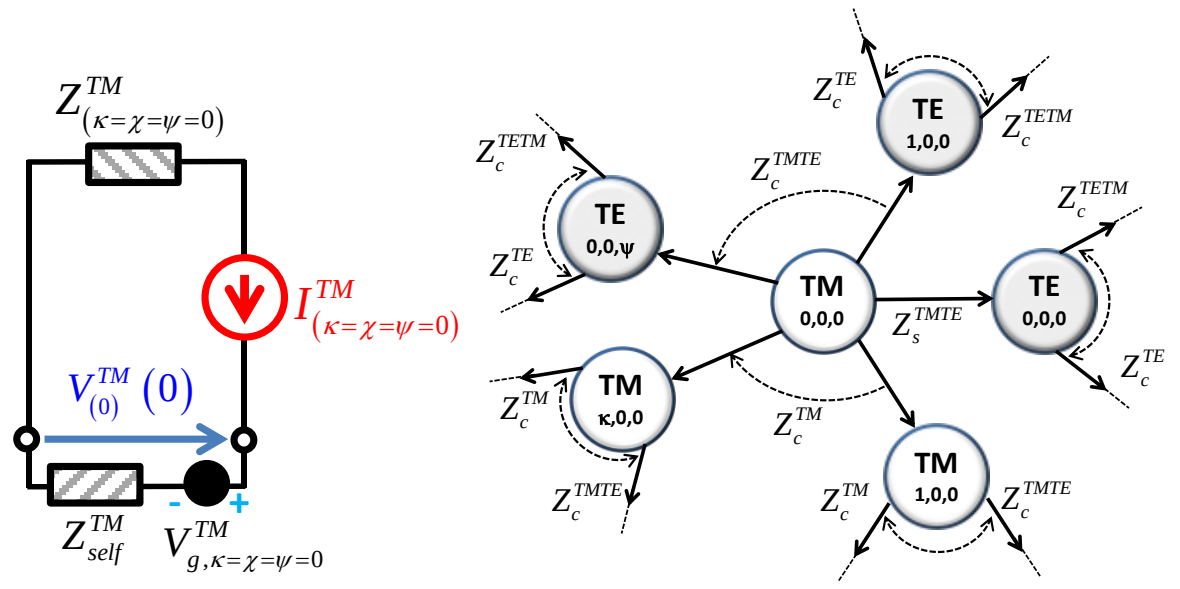

(a) (b)

Figure 15: (a) Excitation of a single mode. (b) General interaction graph for a modulated tensor impedance condition.

The problem defined by (38) in general is not periodic, since tensor components could have different periods. However, it is easy to show that the electromagnetic field can still be expressed as a discrete sum of modes (discrete spectrum).

Let us assume to excite a single mode as shown in Fig. 15a. The energy will *propagate* from the excited mode to other modes by discrete jumps through relation (43) (see Fig. 15b). The energy can reach only modes with wavenumber given by linear combination of $2\pi/p_{\nu,\kappa}$ with integer parameters (representing the number of jumps). This latter aspect assures that the excited modes belong to a countably infinite set. Thus, our problem can still be interpreted as composed by an infinite discrete number of independent transmission lines as for the scalar case ( Fig.2 ). The general transmission line propagation constant can be parametrized as

$$k_z^{(\kappa,\chi,\psi)} = \sqrt{k^2 - \left(k_x^{(0,0,0)} + \kappa\frac{2\pi}{p_{TM}} + \chi\frac{2\pi}{p_{TE}} + \psi\frac{2\pi}{p_{TM,TE}}\right)^2} \tag{53}$$

where $\kappa,\chi,\psi \in \mathbb{N}$, and $\kappa = \chi = \psi = 0$ corresponds to the wavenumber closer to the unmodulated surface one.

As a final remark, we note that the spectrum is *generated* by discrete jumps between nodes, and at each jump the voltage amplitude is attenuated by a factor proportional to *M*. Thus, as for the scalar case the effective impedance of a general transmission line will converge after some iteration.

An alternative formulation can be obtained imposing the resonance condition (14) as a function of the circuits currents, leading to:

$$\begin{cases} I^{TM}_{n,m,l} A^{TM}_{n,m,l} + Z^{TM,TE}_{self} I^{TE}_{n,m,l} + \sum_{\substack{\psi=-\infty \\ \psi \neq l}}^{\infty} Z^{TM,TE}_{c,-\psi} I^{TE}_{n,m,l-\psi} + \sum_{\substack{\kappa=-\infty \\ \kappa \neq n}}^{\infty} Z^{TM}_{c,-\kappa} I^{TM}_{n+\kappa,m,l} = 0 \\ I^{TE}_{n,m,l} A^{TE}_{n,m,l} + Z^{TM,TE}_{self} I^{TM}_{n,m,l} + \sum_{\substack{\psi=-\infty \\ \psi \neq l}}^{\infty} Z^{TM,TE}_{c,-\psi} I^{TM}_{n,m,l-\psi} + \sum_{\substack{\chi=-\infty \\ \chi \neq m}}^{\infty} Z^{TE}_{c,-\chi} I^{TE}_{n,m+\chi,l} = 0 \end{cases} \quad (54)$$

where $A^{\nu}_{n,m,l} = Z^{\nu}_{n,m,l} + Z^{\nu}_{self}$. After writing (54) in matrix form, solutions are found imposing the vanishing of the determinant.

*1) Sinusoidal modulation*

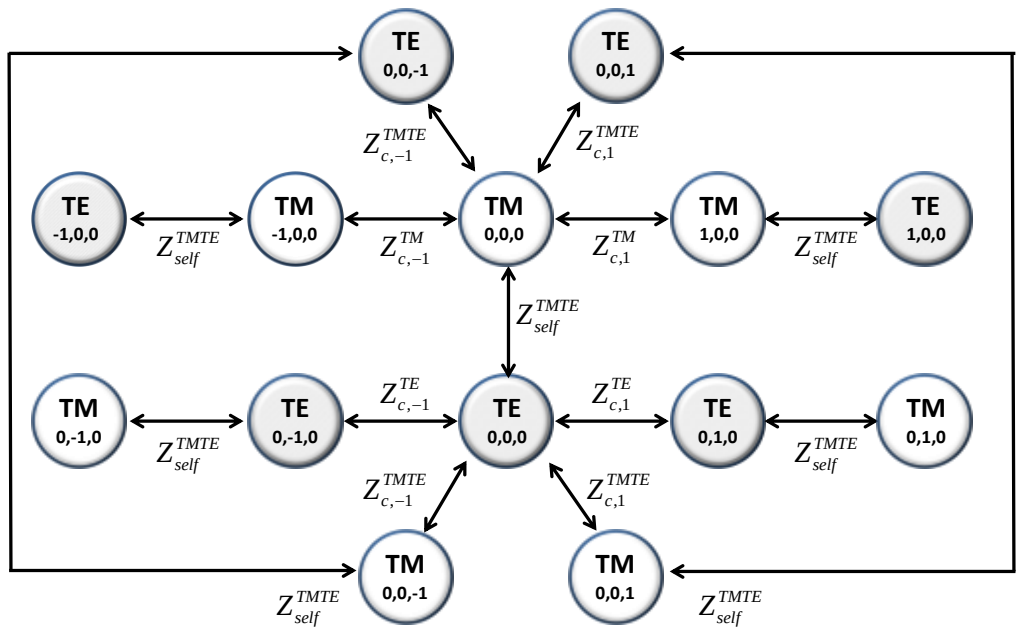

Figure 16: First order interaction graph for a sinusoidally modulated tensorial surface impedance.

As an example, a sinusoidal modulation as that used in [12] is considered. The impedance is defined by the following tensor components:

$$\begin{aligned} Z^{TM,TM} &= j\bar{X}^{TM}_{s}\left(1+M^{TM,TM}\cos\left(2\pi x/p_{TM}+\varphi_{TM}\right)\right) \\ Z^{TM,TE} &= j\bar{X}^{TM,TE}_{s}\left(1+M^{TM,TE}\cos\left(2\pi x/p_{TM,TE}+\varphi_{TM,TE}\right)\right) \\ Z^{TE,TE} &= j\bar{X}^{TE}_{s}\left(1+M^{TE}\cos\left(2\pi x/p_{TE}+\varphi_{TE}\right)\right) \end{aligned} \quad (55)$$

where $p_{TM} \neq p_{TE} \neq p_{TM,TE}$. Using (55) in (42) leads to

$$Z^{\nu}_{c,n} = \begin{cases} jX^{\nu}_{s} M^{\nu} e^{jn\varphi_{\nu}}/2 & \text{if } n = \pm 1 \\ 0 & \text{otherwise} \end{cases}. \quad (56)$$

Eq.(56) states that each node (mode) is related with two others modes with the same polarizations through $Z^{TM}_{c,\pm1}$ and with three modes of the other polarization state through $Z^{TM,TE}_{self}$, $Z^{TM,TE}_{c,\pm1}$.

The first order interaction graph is depicted in Fig. 16. The corresponding effective impedance for $TM_{0,0,0}$ mode can be written as:

$$\begin{aligned} Z^{TM}_{Eff(0,0,0)}\left(k_{\rho}\right) &= \left(Z^{TM}_{c}\right)^{2}\left(\vartheta^{TE,TM}_{-1a}+\vartheta^{TE,TM}_{1a}\right)+Z^{TM,TE}_{self}A^{TM,TE}+Z^{TM}_{self} \\ &+\left(Z^{TM,TE}_{c}\right)^{2}\left[\vartheta^{TM,TE}_{-1c}+\vartheta^{TM,TE}_{1c}-A^{TETM}\left(\xi_{-1c}-\xi_{1c}\right)\right] \end{aligned} \quad (57)$$

where $\vartheta^{\nu,w}_{\kappa}$ and $A^{TM,TE}$ are defined in appendix A.

The propagation constant is then obtained using (57) in (22).

Fig. 17 shows the real and imaginary part of the propagation constant as a function of the three modulation indexes for a tensor surface impedance with average impedances $X^{TM}_{s} = 400$, $X^{TE}_{s} = 300$, $X^{TMTE}_{s} = -100$, and modulation given by $\sin(\theta_{TM}) = 0.1$, $\sin(\theta_{TE}) = 0.2$, $\sin(\theta_{TMTE}) = 0.3$.

Continuous lines are relevant to the exact solution obtained with (54), while the dashed lines are relevant to solutions obtained by (22). As for the previous example, since the average impedance support a dominant TM current, higher variations of the propagating constant are obtained with modulations of the TM tensorial component.

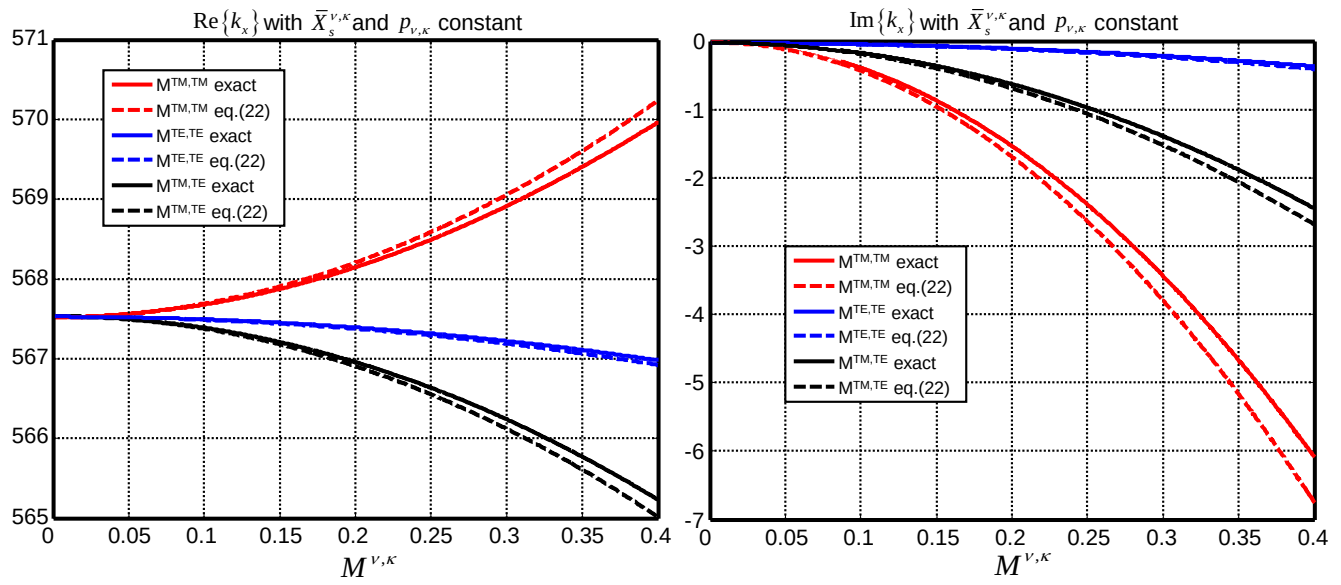

Figure 17: Real and imaginary part of the propagation constant over a tensor impedance surface with parameters: $X^{TM}_{s} = 400$, $X^{TE}_{s} = 300$, $X^{TM,TE}_{s} = -100$, $\sin(\theta_{TM}) = 0.1$, $\sin(\theta_{TE}) = 0.2$, $\sin(\theta_{TMTE}) = 0.3$.

## *C. General modulation*

In this section we consider a modulated impedance tensor in a general orthogonal framework, namely

$$\underline{\underline{\mathbf{Z}}}'_{s} = \begin{bmatrix} Z'_{1,1} & Z'_{1,2} \\ Z'_{2,1} & Z'_{2,2} \end{bmatrix} \quad (58)$$

where each component has the form (40) where now $\nu,\kappa$ =1/2 and equations (41)-(43) are still valid.

It is convenient to express eq. (58) in the canonical TM/TE basis:

$$\underline{\underline{\mathbf{Z}}}_{s} = \begin{bmatrix} Z_{TM,TM} & Z_{TM,TE} \\ Z_{TE,TM} & Z_{TE,TE} \end{bmatrix} = \underline{\underline{\mathbf{R}}}^{T}\cdot\underline{\underline{\mathbf{Z}}}'_{s}\cdot\underline{\underline{\mathbf{R}}} \quad (59)$$

where $\mathbf{R}$ is the orthogonal transformation matrix from the canonical framework into the general one.

From (59) it follows that each component of the $\mathbf{Z}_{s}$ matrix is a linear combination of the $\mathbf{Z}'$ matrix elements:

$$Z^{\nu,\kappa}(x) = \alpha^{\nu,\kappa}_{1,1}Z'^{1,1} + \alpha^{\nu,\kappa}_{1,2}Z'^{1,2} + \alpha^{\nu,\kappa}_{2,1}Z'^{2,1} + \alpha^{\nu,\kappa}_{2,2}Z'^{2,2} \quad (60)$$

where $\alpha^{\nu,\kappa}_{n,m}$ are given in appendix B.

Eq. (60) can be recast in a form similar to (42):

$$\begin{aligned} Z^{\nu,\kappa}(x) = Z^{\nu,\kappa}_{self} + \sum_{m=-\infty}^{\infty} \Bigg\{ & Z^{\nu,\kappa}_{c11,m} e^{-jm\frac{2\pi}{p_{1,1}}x} + Z^{\nu,\kappa}_{c12,m} e^{-jm\frac{2\pi}{p_{1,2}}x} \\ & + Z^{\nu,\kappa}_{c21,m} e^{-jm\frac{2\pi}{p_{2,1}}x} + Z^{\nu,\kappa}_{c22,m} e^{-jm\frac{2\pi}{p_{2,2}}x} \Bigg\} \end{aligned} \quad (61)$$

where

$$\begin{aligned} Z^{\nu,\kappa}_{self} &= \alpha^{\nu,\kappa}_{1,1} j\bar{X}^{1,1}_{s} + \alpha^{\nu,\kappa}_{1,2} j\bar{X}^{1,2}_{s} + \alpha^{\nu,\kappa}_{2,1} j\bar{X}^{2,1}_{s} + \alpha^{\nu,\kappa}_{2,2} j\bar{X}^{2,2}_{s} \\ Z^{\nu,\kappa}_{c,pq,m} &= \alpha^{\nu,\kappa}_{1,1} j\bar{X}^{p,q}_{s} M^{p,q} c^{p,q}_{n} \end{aligned} \quad (62)$$

and $p,q=1/2$. Now, the effective impedance and the dispersion relation can be calculated using in the proposed circuital approach using the impedances defined in (62).
The alternative matrix formulation can be obtained imposing the resonance condition (14) as a function of the circuit currents, leading to a matrix formulation.
The same procedure can be also applied to modulating functions presenting a discrete Fourier spectrum.

## IV. TRANSPARENT IMPEDANCE MODULATION

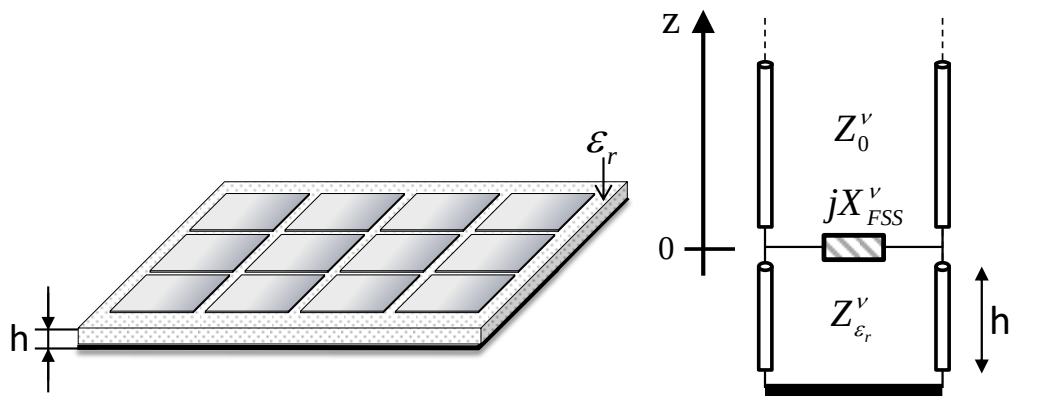

(a) (b)

Figure 18. (a) Surface impedance composed by periodic metallic elements printed over a grounded substrate. (b) Equivalent circuit.

In metasurface applications, surface impedances are obtained by printing periodic small patches on a grounded dielectric slab as shown in Fig. 18a [2]. Figure 18b shows the equivalent transmission line for the TE/TM polarization. The dielectric slab of height $h$ and relative permittivity $\varepsilon_r$ is represented by a piece of transmission line with characteristic impedance $Z_{\varepsilon_r}^\nu$, while the printed elements are described by the parallel impedance $jX_{FSS}^\nu$ ( homogenized scalar boundary condition). Changing the local geometry of the patches allows modulating the corresponding parallel impedance [10].

The impedance at *z=0* can be obtained using transmission line theory as

$$Z_s^{TM} = \frac{jZ_{FSS}^{TM} Z_{Lslab}^{TM} \tan\left(k_{z_1} h\right)}{Z_{FSS}^{TM} + jZ_{Lslab}^{TM} \tan\left(k_{z_1} h\right)}. \qquad (63)$$

If $Z_{FSS}^{TM}$ is periodically modulated, then also $Z_s^{TM}$ is a periodic function. Thus, the propagation characteristics of waves guided by these structures can be analyzed with the proposed method. As a final remark, we would like to point out the fact that a sinusoidal modulation of $jX_{FSS}^\nu$ as in [10] does not correspond to a sinusoidal modulation of the impedance at *z=0*. As for the general case, all the Fourier harmonics describing (63) have to be considered.

## V. CONCLUSION

A procedure to derive the propagation characteristics and field distributions of waves guided by scalar and tensorial reactance surfaces modulated by periodic or discrete Fourier spectrum functions in the propagation direction was presented. The technique is based on an equivalent system of coupled lumped circuits. The complex propagation constant can be determined via the introduction of an effective impedance or equivalently through an infinite matrix equation. Examples of scalar and tensorial modulation were presented and their connection with metasurface antenna applications discussed.

## APPENDIX A

This appendix presents the definition of the quantities used in section III.B.1:

$$A^{TM,TE} = \frac{\left(Z_c^{TM,TE}\right)^2\left(\xi_{-1c}+\xi_{1c}\right) - Z_{self}^{TETM}}{B^{TE}},$$

$$B^{TE} = Z_{(0)}^{TE} + Z_{self}^{TE} + \left(Z_c^{TE}\right)^2\left(\vartheta_{-1c}^{TM,TE}+\vartheta_{1c}^{TM,TE}\right) + \left(Z_c^{TMTE}\right)^2\left(\vartheta_{-1c}^{TE,TM}+\vartheta_{1c}^{TE,TM}\right),$$

$$\vartheta_\kappa^{\nu,w} = \frac{\left(Z_\kappa^\nu + Z_{self}^\nu\right)}{\left(Z_{self}^{TMTE}\right)^2 - \left(Z_\kappa^w + Z_{self}^w\right)\left(Z_\kappa^\nu + Z_{self}^\nu\right)},$$

$$\xi_\kappa = \frac{Z_{self}^{TMTE}}{\left(Z_{self}^{TMTE}\right)^2 - \left(Z_\kappa^{TE} + Z_{self}^{TE}\right)\left(Z_\kappa^{TM} + Z_{self}^{TM}\right)}.$$

## APPENDIX B

This appendix reports the calculation of the constants used in section III.C.
The TM/TE electric field transverse unit vectors can be written in terms of the tangent propagation vector $\hat{\mathbf{k}}_t$ as

$$\begin{aligned} \hat{\mathbf{u}}_t^{TM} &= \hat{\mathbf{k}}_t = \hat{\mathbf{x}} \\ \hat{\mathbf{u}}_t^{TE} &= \hat{\boldsymbol{\alpha}}_t = \hat{\mathbf{n}} \times \hat{\mathbf{k}}_t = \hat{\mathbf{y}} \end{aligned}. \qquad (B1)$$

Thus, the rotation matrix from the canonical framework to a general one defined by the unit vectors: $\hat{\mathbf{u}}_1 = u_{1x}\hat{\mathbf{x}} + u_{1y}\hat{\mathbf{y}}$, $\hat{\mathbf{u}}_2 = u_{2x}\hat{\mathbf{x}} + u_{2y}\hat{\mathbf{y}}$, is given by

$$\underline{\underline{\mathbf{R}}} = \begin{bmatrix} \hat{\mathbf{k}}_t \cdot \hat{\mathbf{u}}_1 & \hat{\boldsymbol{\alpha}}_t \cdot \hat{\mathbf{u}}_1 \\ \hat{\mathbf{k}}_t \cdot \hat{\mathbf{u}}_2 & \hat{\boldsymbol{\alpha}}_t \cdot \hat{\mathbf{u}}_2 \end{bmatrix} = \begin{bmatrix} u_{1x} & u_{1y} \\ u_{2x} & u_{2y} \end{bmatrix}. \qquad (B2)$$

Finally, from $\underline{\underline{\mathbf{Z}}}_s = \underline{\underline{\mathbf{R}}}^T \cdot \underline{\underline{\mathbf{Z}}}_s' \cdot \underline{\underline{\mathbf{R}}}$ and (A2) it follows:

$$\begin{aligned} \alpha_{n,\mathrm{m}}^{TM,TM} &= u_{\mathrm{nx}} u_{mx} \\ \alpha_{\mathrm{n,m}}^{TM,TE} &= u_{\mathrm{nx}} u_{\mathrm{my}} \\ \alpha_{\mathrm{n,m}}^{TE,TM} &= u_{ny} u_{mx} \\ \alpha_{n,m}^{TE,TE} &= u_{ny} u_{my} \end{aligned} \qquad (B3)$$

where $n,m = 1,2$.

## ACKNOWLEDGMENT

The author wish to thank G. Valerio of Sorbonne Universités UPMC, for many stimulating discussions on the subject of periodic structures.